\documentclass[fleqn,11pt,twoside]{article}

\usepackage{amsthm,amsthm,amssymb, color, xcolor,epsfig, graphics, subfigure}

\usepackage{amsmath, graphicx, latexsym, lscape}

\usepackage{amsfonts,verbatim,xkeyval,bm, amsthm, upgreek, amscd, wasysym, amssymb}
\usepackage{hyperref}

\makeatletter
\newcommand{\copyrightnote}[2]{{\renewcommand{\thefootnote}{}
 \footnotetext{\small\it
\begin{flushleft}
 \copyright \ #1   #2  
\end{flushleft}}}}

\newcommand{\Name}[1]{\begin{flushleft}
                       \LARGE \bf #1
                       \end{flushleft}\vspace{-3mm}}

\newcommand{\Author}[1]{\begin{flushleft}
                       \it #1 \end{flushleft}}

\newcommand{\Address}[1]{\begin{flushleft}
                       \it #1 \end{flushleft}}

\newcommand{\Date}[1]{\begin{flushleft}
                      \small  \it #1 \end{flushleft}}

\newcommand{\evenhead}{Author \ name}
\newcommand{\oddhead}{Article \ name}

\renewcommand{\@evenhead}{
\hspace*{-3pt}\raisebox{-15pt}[\headheight][0pt]{\vbox{\hbox to \textwidth
{\thepage \hfil \evenhead}\vskip4pt \hrule}}}
\renewcommand{\@oddhead}{
\hspace*{-3pt}\raisebox{-15pt}[\headheight][0pt]{\vbox{\hbox to \textwidth
{\oddhead \hfil \thepage}\vskip4pt\hrule}}}
\renewcommand{\@evenfoot}{}
\renewcommand{\@oddfoot}{}

\long\def\@makecaption#1#2{%
  \vskip\abovecaptionskip
  \sbox\@tempboxa{\small \textbf{#1.}\ \ #2}%
  \ifdim \wd\@tempboxa >\hsize
    {\small \textbf{#1.}\ \ #2}\par
  \else
    \global \@minipagefalse
    \hb@xt@\hsize{\hfil\box\@tempboxa\hfil}%
  \fi
  \vskip\belowcaptionskip}

\newcommand{\JNMPnumberwithin}[3][\arabic]{%
  \@ifundefined{c@#2}{\@nocounterr{#2}}{%
    \@ifundefined{c@#3}{\@nocnterr{#3}}{%
      \@addtoreset{#2}{#3}%
      \@xp\xdef\csname the#2\endcsname{%
        \@xp\@nx\csname the#3\endcsname .\@nx#1{#2}}}}%
}

\newcommand{\resetfootnoterule} {
  \renewcommand\footnoterule{%
  \kern-3\p@
  \hrule\@width.4\columnwidth
  \kern2.6\p@}
}

\renewcommand{\footnoterule}{}

\makeatother

\theoremstyle{definition}

\begin{document}

\renewcommand{\evenhead}{ {\LARGE\textcolor{blue!10!black!40!green}{{\sf \ \ \ ]ocnmp[}}}\strut\hfill 
T Ekelchik and A Marchesiello
}
\renewcommand{\oddhead}{ {\LARGE\textcolor{blue!10!black!40!green}{{\sf ]ocnmp[}}}\ \ \ \ \   
Superintegrable 2D systems in magnetic fields 
}

\thispagestyle{empty}
\newcommand{\FistPageHead}[3]{
\begin{flushleft}
\raisebox{8mm}[0pt][0pt]
{\footnotesize \sf
\parbox{150mm}{{Open Communications in Nonlinear Mathematical Physics}\ \  \ {\LARGE\textcolor{blue!10!black!40!green}{]ocnmp[}}
\ \ Vol.6 (2026) pp
#2\hfill {\sc #3}}}\vspace{-13mm}
\end{flushleft}}

\FistPageHead{1}{\pageref{firstpage}--\pageref{lastpage}}{ \ \ Article}

\strut\hfill

\strut\hfill

\copyrightnote{The author(s). Distributed under a Creative Commons Attribution 4.0 International License}

\Name{Superintegrable 2D systems in magnetic fields with a parabolic type integral}

\Author{Tatiana Ekelchik$^1$, Antonella Marchesiello$^2$}

\Address{ Faculty of Information Technology, 
Czech Technical University in Prague, Th\' akurova 9, 160 00 Prague \\
e-mail: $^1$ekelctat@fit.cvut.cz, $^2$marchant@fit.cvut.cz }

\Date{Received May 29, 2026; Accepted July 5, 2026}

\setcounter{equation}{0}

\smallskip

\noindent
{\bf Citation format for this Article:}\newline
T Ekelchik and A Marchesiello, 
Superintegrable 2D systems in magnetic fields with a parabolic type integral,
{\it Open Commun. Nonlinear Math. Phys.}, {\bf 6}, ocnmp:18305,
 \pageref{firstpage}--\pageref{lastpage}, 2026.

\strut\hfil

\noindent
{\bf The permanent Digital Object Identifier (DOI) for this Article:}\newline
{\it 10.46298/ocnmp.18305}

\strut\hfill

\begin{abstract}
\noindent 
We consider the problem on the existence of two dimensional superintegrable systems in the presence of a magnetic field in the two dimensional Euclidean space. We assume the existence of two integrals of motion, besides the Hamiltonian, that are quadratic polynomials in the momenta. This problem was already studied in the cases where one integral is of Cartesian or polar type \cite{BeWin}. We continue the investigation by assuming that one of the integrals is of parabolic type and the second integral is of elliptic or (``non-standard'') parabolic type, confirming so far that, on the Euclidean plane, the only two dimensional superintegrable system with quadratic integrals is the one with constant magnetic field and constant electrostatic potential.
\end{abstract}

\label{firstpage}


\section{Introduction}

In recent years, the search for new integrable and superintegrable systems has been the focus of an intense research activity \cite{MSW,MSW2,Marchesiello2022,EscobarRuiz2020,Yurdusen2025,Nikitin2024,KScomplex2023,delOlmo2021,Snobl2025,MarquetteParr2025, Hamilton2024}. Several new (super)integrable systems have been discovered,  including e.g. integrable quantum dots   \cite{Dun2024}, superintegrable helical undulators \cite{KMS2022} and superintegrable magnetic monopoles \cite{Marchesiello2025, MRS}.  However, despite this extensive investigation, a complete classification of integrable and superintegrable systems in the presence of magnetic fields, even in the simplest case  of quadratic integrals (i.e., the case where the integrals are quadratic polynomials in the momenta), is still not achieved.

Let us recall that a classical Hamiltonian system in $n$ degrees of freedom which does not explicitly depend on time is said to be integrable if it admits 
$n$ functionally independent constants of motion that are pairwise in involution, i.e., $n$ integrals that Poisson commute
with the Hamiltonian and pairwise with each other. For superintegrability we require additional independent integral(s) (which cannot be in involution with all the others).

The classification of (super)integrable systems in the Euclidean space is completed for ``natural'' Hamiltonian systems (i.e. systems in which the Hamiltonian consists only of kinetic and potential energy) in the case where the integrals are at most second order polynomials in the momenta in two and three dimensions \cite{Makarov1967,Snobl2025,Evans,Miller2013}.
 
However, the classification problem is still open in the case also linear terms in the velocity are present in the Hamiltonian, due e.g. to the presence of a magnetic field. In this case, only partial results on the conditions for the existence of quadratic integrals are known in two \cite{HietarintaReview,CharHuWin, BeWin, DoGraRaWin,Pucacco2004,Pucacco_2005, Hamilton2024}  and three \cite{MSW,MSW2,Kubu_2021,KScomplex2023, BertrandSnobl,Marchesiello2019,Marchesiello2022} dimensions.  However, a complete classification is still not achieved, even in the simpler two-dimensional (2D) case with second order integrals. Not to mention that systems with more general type of integrals, such as higher order polynomials in the momenta, rational or transcendental integrals can exist \cite{HietarintaReview}.
 
The aim of this work is to contribute to the solution of the 2D classification problem. This problem has already been addressed and solved on the 2D Euclidean space in the case one of the integrals is quadratic of Cartesian or polar type \cite{BeWin}, both in the classical and quantum case. These integrals are so called because in the absence of a magnetic field the system would separate in Cartesian and polar coordinates, respectively. Several quadratically integrable systems were found in this case.  However, let us notice that already in two dimensions, in presence of a magnetic field quadratic integrability does not imply separability \cite{HietarintaReview, BeWin, Benenti_2001}.

With the assumption of \cite{BeWin},  the only  quadratically superintegrable system found was the system with constant magnetic field and constant electrostatic potential.  Since the electrostatic potential can be set to zero in this case, in the following we often refer to this system shortly as the CMF (constant magnetic field) system.

Here we continue in the investigation by assuming one of the two integrals to be of parabolic type (thus, the system would be separable  in parabolic coordinates for a vanishing magnetic field). In this case, even the classification of  2D integrable systems is not yet achieved, although some results already exist in this direction \cite{ Pucacco_2005, Hamilton2024}. In this manuscript, we contribute to the classification of 2D quadratically superintegrable systems with one parabolic type integral in the classical case. We assume the additional integral to be of elliptic type (i.e. the system would separate in the elliptic coordinates in the absence of magnetic field) or of ``non-standard'' parabolic type. By ``non-standard'' we mean that this additional integral could be transformed to a parabolic type integral by Euclidean transformations, but then these would affect the form of the other parabolic integral, of course.  Systems with more general structure of the quadratic terms in the integrals, not related to an orthogonal coordinate system, are known to exist in three dimensions \cite{Marchesiello2022,KMS2022}, even without magnetic field \cite{Snobl2025}.

Since the CMF system, in proper gauge choice, has actually three first order integrals, given by the two linear momenta $p_1,p_2$ and the third component of the angular momenta $\ell_3$ \cite{BeWin}, it will clearly possess also a parabolic type integral \cite{DoGraRaWin}. However, it remains open the question whether this is the only superintegrable system with a parabolic type integral. Indeed,  to the best of our knowledge, in the presence of a magnetic field, the only known superintegrable system on the plane, with integrals that are at most quadratic, is the CMF system. Thus, a more interesting open question is whether the CMF system is the only quadratically superintegrable system in the 2D Euclidean space (in the presence of a magnetic field).

In the absence of a magnetic field, integrals related to separation in subgroup type coordinates, such as Cartesian and polar coordinates, correspond to richer results concerning (super)integrability with respect to integrals related to non-subgroup type coordinates \cite{Miller2013}. Such a disparity has been particularly observed  when  higher order integrals are considered \cite{Marchesiello2015a, TW, PPW, Gravel, Miller2013}. Therefore, on one hand, it would be reasonable to expect that, since only one quadratically superintegrable system exists when subgroup type integrals are considered, no other 2D quadratically superintegrable system can be found, with non-vanishing magnetic field, when considering non-subgroup type integrals (that is, parabolic and elliptic type integrals).
However, on the other hand, the free motion in Euclidean plane allows maximal superintegrability, with three first order integral given by $p_1,p_2,\ell_3$. These correspond to  the generators of the 2D Euclidean algebra and it is known that every integral polynomial in the momenta must have the leading order terms polynomial in $p_1,p_2,\ell_3$ \cite{Miller2013}. Therefore, we can not exclude a priori the existence of a  superintegrable system with  non-subgroup type integrals that would trivialize to  $p_1,p_2$ or $\ell_3$  in the limit for vanishing magnetic field and vanishing electrostatic potential. 

Here we started  to look for an answer to this question by assuming the existence of two quadratic integrals,  one of which of parabolic type.  The results we obtained so far confirm that the only 2D superintegrable system, in presence of a magnetic field, is the CMF system.

The manuscript is organized as follows. In the following Section \ref{sec:system}, we introduce the system under study and fix the notation.  The  structure of the integrals we are considering is presented in Section \ref{sec:goal}. In Section \ref{sec:strategy} we describe our approach to the solution of the determining equations of the integrals, that are given explicitly in Sections \ref{sec:eqX1} and \ref{sec:eqX2}, together with their compatibility conditions. We start by solving such equations in the case where, besides the parabolic integral,  there also exists an elliptic type integral, see Section \ref{sec:par-ell-Sol}. In the following Sections \ref{sec:Par-Par}-\ref{sec:nonstandardPar}, we consider the case where also the second integral is of parabolic or, more generally, of non-standard parabolic type.  Since the solution with constant magnetic field is recurrent in many branches of the computation, we consider this case separately in Section \ref{sec:ConstB}.  Conclusions and future perspectives are discussed in Section \ref{Sec:conclusions}. 

\section{The system}\label{sec:system}
We consider the Hamiltonian describing the motion of a charged particle on the plane in the presence of a magnetic field. Namely, by choosing units so that the particle has mass $1$ and charge $-1$, we consider a Hamiltonian  of the form
\begin{equation}\label{HamiltonianV}
    H = \frac {1} {2} (p_1^2 + p_2^2) + A_1 (x, y) p_1 + A_2(x, y)p_2+ 
  V (x, y),
\end{equation}
where $(x,y)$ are the space variable, $p_1,p_2$ the corresponding momenta, $V(x,y)$ is the scalar potential and $A(x,y)=(A_1(x,y),A_2(x,y))$ is the vector potential. 


The magnetic field $\vec {\mathcal B}$ is assumed to be orthogonal to the $(x,y)$ plane and to depend only on the space variables. Thus,
\begin{equation}\label{magField}
    \vec{\mathcal B}(x,y) = (0,0,B(x,y))\,,
\end{equation}
where
\begin{equation}
B(x,y)= \partial_x A_2(x,y) - \partial_y A_1(x,y).
 \end{equation}
For convenience, let us introduce the covariant momenta
\begin{equation}
    P_1^A = p_1 + A_1(x,y), \quad P_2^A = p_2 + A_2(x,y).
\end{equation}
In gauge invariant form, the Hamiltonian \eqref{HamiltonianV} then reads
\begin{equation}\label{HamiltonianW}
    H = \frac {1} {2} ((P_1^A)^2 + (P_2^A)^2)+  W(x, y)\,,
\end{equation}
where 
\begin{equation}
    W=V-\frac{1}{2}(A_1^2+A_2^2)\label{potW}
\end{equation}
is the so called electrostatic potential.

\section{The integrals}\label{sec:goal}

The aim of this work is to study the conditions on the magnetic field \eqref{magField} and electrostatic potential $W(x,y)$, cf. \eqref{potW}, so that the system \eqref{HamiltonianW} is quadratically superintegrable, i.e. it admits two  independent integrals $X_1, X_2$ that are second order polynomials in the momenta. 

The highest order terms in the integrals are known to be polynomials in $p_1, p_2, l_3$ with constant coefficients \cite{Miller2013}. Here and in the following, let $l_3$ denote the third component of the angular momentum and
\begin{equation}
    L_3^A = x P_2^A - y P_1^A
\end{equation}
the covariant angular momentum. 

By Euclidean transformations, it is always possible to simplify the  structure of one of the integrals, let us say $X_1$ to one of the following forms \cite{DoGraRaWin}:
\begin{itemize}
\item[i)] $(P_1^A)^2+ \ldots$,
\item[ii)] $(L_3^A)^2+ \ldots$,
\item[iii)] $P_2^A L_3^A + \ldots$,
\item[iv)] $(L_3^A)^2+ a ((P_1^A)^2- (P_2^A)^2)  + \ldots$,
\end{itemize}
where ``$\ldots$'' denotes lower order terms in the momenta, $a$ is an arbitrary constant. In the absence of a magnetic field, each of the integrals listed above corresponds to separation of the system in one orthogonal coordinate system of  the Euclidean plane, namely Cartesian (i), polar (ii), parabolic (iii) and elliptic coordinates (iv). Correspondingly, the integrals are said to be of Cartesian, polar, parabolic and elliptic type.

The problem of the existence of an additional independent quadratic integral $X_2$ for the system \eqref{HamiltonianW} has already been studied and solved in the case one of the integrals is of Cartesian or polar type  \cite{BeWin}. We focus here on the case in which the integral $X_1$ is of parabolic type, namely
\begin{equation}\label{IntX1}
    X_1=L_3^AP_2^A+k_1(x,y)P_1^A+k_2(x,y)P_2^A+m_1(x,y),
\end{equation} 
where $k_1, k_2$ and $m_1$ are smooth functions.
We assume the second integral $X_2$ to be quadratic in the momenta as well. Once the form of the integral $X_1$ is chosen, it is not possible to use Euclidean transformations to also simplify $X_2$ to one of the types i)-iv) without affecting $X_1$. Thus, $X_2$ must be taken in the most general form  for a quadratic integral, given by (modulo a subtraction of the Hamiltonian)
\begin{eqnarray}
    X_2&=&a((P_1^A)^2-(P_2^A)^2)+bP_1^AP_2^A+c(L_3^A)^2+dL_3^AP_1^A \nonumber\\
    &+&s_1(x,y)P_1^A+s_2(x,y)P_2^A+m_2(x,y)\label{IntX2},
\end{eqnarray}
where $s_1,s_2, m_2$ are smooth functions, $a,b,c,d$ $\in\mathbb R$. 
In order to be integrals for  the system \eqref{HamiltonianW},   $X_{1}$ and $X_{2}$ have to satisfy the following commutativity condition with Hamiltonian:
\begin{equation}\label{CondInt}
\{H, X_{1}\} = 0, 
\qquad 
\{H, X_{2}\} = 0,
\end{equation}
where $\{\cdot,\cdot \}$ is the  Poisson bracket.
Conditions \eqref{CondInt} provide the so called determining equations for the integrals. We will write them explicitly for both integrals in Sections \ref{sec:eqX1}, \ref{sec:eqX2}. 

In the absence of a magnetic field, the solution of equations \eqref{CondInt} is known.  In this case, a 2D quadratically superintegrable system is multi-separable, that is, separable in more than one coordinate system on the plane. The converse is also true: a multi-separable system is quadratically superintegrable. As a consequence, both integrals must be of one of the forms i)-iv)  listed above \cite{Miller2013, Evans}.  
However, in the presence of a magnetic field, the relationship between the existence of a quadratic integral and the separability of the system (in configuration space) no longer holds \cite{Shapovalov1972, Benenti_2001}. 
Furthermore, even if in the absence of  magnetic field the only superintegrable systems with non-vanishing potential and integral of the form \eqref{IntX1}  have the second integral related to subgroup-type coordinates (i.e. Cartesian and polar) \cite{Miller2013}, we can not exclude a priori the existence of a superintegrable system with integrals $X_1$ and $X_2$ of the more general form \eqref{IntX2} for non-vanishing magnetic field. It could be e.g. a system that  in the limit for vanishing magnetic field would trivialize to free motion.  Or we could find a magnetic field depending on the constants $a,b,c,d$  in such a way that in the limit for vanishing magnetic field the integral $X_2$ would reduce to subgroup type integral.  (Super)integrable systems with so called non-standard or generalized quadratic integrals, that do not correspond to separation of variables (in configuration space), are known to exist  \cite{Marchesiello2022,KMS2022} in 3D, even without magnetic field \cite{Snobl2025}.
Thus, there is the need to study the conditions for the existence of a pair of integrals as in \eqref{IntX1}, \eqref{IntX2}, in order to advance in the solution of the classification problem of 2D superintegrable systems with non-vanishing magnetic field.

\section{Approach to the problem}\label{sec:strategy}
 A ``brute force'' approach to solve the determining equations for the integrals, cf. \eqref{CondInt}, clearly fails, due to the number of unknowns (the magnetic field $B$, the potential $W$, the functions $k_j,s_j$, $m_j$ and the constants $a,b,c,d$ in the integrals). In fact, to our knowledge, considering the integrability problem on the existence of  one quadratic integral $X_1$ of parabolic type, it was not possible so far to solve the determining equations, although partial results exist in the literature \cite{Pucacco_2005, Hamilton2024}.

However, by imposing the existence of a second integral $X_2$, we further restrict the equations for the integral $X_1$. This allows then for the solution of the determining equations for both integrals $X_1$ and $X_2$, at least in some special cases, yielding some answers on the existence of 2D quadratically superintegrable systems with a parabolic type integral.

To achieve a solution of the determining equations \eqref{CondInt}, it will be effective in the following to first derive compatibility conditions for them and start by solving such conditions. We shall start from the compatibility conditions for the magnetic field. However, before proceeding, some observations are in order.
\begin{enumerate}
\item We shall exclude the case in which all constants $a,b,c,d$ are zero, since otherwise the system would have one first order integral.  In this case, it was proven that the only superintegrable system is the one with constant magnetic field and constant electrostatic potential \cite{BeWin}. In proper gauge choice, the independent integrals of such system are $\ell_3$ and one linear momentum. Thus, the system with constant potential and constant magnetic field trivially also admits  integrals of the type $X_1$ and $X_2$ for arbitrary values of the constants $a,b,c,d$ and vanishing  $k_j, s_j$ and $m_j$.
\item For  $b=c=d=0$ and $a=b=d=0$ the integral $X_2$ simplifies to Cartesian (after adding $2 a H$) and polar type integral, respectively.  As we have already observed, the superintegrability problem  in these cases is solved in \cite{BeWin}, where it was found that the only system that is quadratically superintegrable (with one Cartesian type or one polar type integral) is the system with constant magnetic field and constant electrostatic potential. 
\item In case $c=d=0$, $b\neq0$  by rotation the integral $X_2$ can be transformed into a Cartesian type integral (the transformation would, of course, also change the integral $X_1$). Thus, we can reduce to the problem already studied in \cite{BeWin}, as well.
\end{enumerate}
In the following, we are going to solve the determining equations \eqref{CondInt} when we are not in cases 1-3  listed above. 
Namely,
for $c\neq0$ we will consider  in the following  $d=b=0$, $a\neq 0$ in order to study the case where $X_2$ is of elliptic type.  We prefer to postpone a complete  study of case $c\neq0$, for arbitrary values of the constants $a,b,d$ for future work and  focus in this manuscript on case $c=0$, $d\neq0$ (in order to exclude all known cases 1-3). This includes the case  $a=b=c=0$ where $X_2$ is a parabolic type integral. For $a,b,d$ not zero, in principle, the integral $X_2$ could be transformed into parabolic type integral by Euclidean transformations (however, this would clearly change the form of the integral $X_1$). Thus, for $c=0$, $d\neq 0$ and at least one constant $a$ or $b$ not zero, we speak of non-standard parabolic type integral $X_2$.

As anticipated before, we shall start from the compatibility conditions for the magnetic field.  Their solution depends on the constants $a,b,c,d$ of the integral $X_2$. 

Once the compatibility conditions for the magnetic field are solved, it is effective to proceed with the compatibility condition for the $s_j$ and $k_j$ functions, given by \eqref{EQksW}. This condition is obtained by combining the determining equations for both integrals together, thus it strongly imposes susperintegrability on the system. By imposing equation  \eqref{EQksW}, we will then always conclude that the only possible superintegrable system is the CMF system, for all values of the constants $a,b,c,d$ that we considered in this work.

\section{Determining equations for the integrals and their compatibility conditions}

Conditions \eqref{CondInt} give the determining equations of the integrals\cite{MSW}, to be solved for the magnetic field $B$, the electrostatic potential $W$ and the functions $k_j, s_j, m_j$. For the sake of completeness, let us derive them below explicitly for both integrals $X_1$ and $X_2$.

\subsection{Determining equations for $X_1$}\label{sec:eqX1}
The conditions \eqref{CondInt} are polynomial in the momenta $p_1, p_2$. Therefore, they are satisfied if and only if the coefficients of all monomials in $p_1, p_2$, collected order by order, equal zero. This provides the determining equations for the integrals. In the following, we derive such equations for the integral $X_1$, cf. \eqref{IntX1}.

\paragraph{Second-order equations:} 
these conditions follow from requiring that the coefficients of $p_1^2$, $p_2^2$, and $p_1p_2$  vanish in \eqref{CondInt} and read
\begin{equation}
    y B(x,y) - \partial_x k_1(x,y) = 0, \label{eqk11}
\end{equation}
\begin{equation}
    -y B(x,y) - \partial_y k_2(x,y) = 0,\label{eqk12}
\end{equation}
\begin{equation}
    2x B(x,y) +\partial_y k_1(x,y) + \partial_x k_2(x,y) = 0.\label{eqk13}
\end{equation}

\paragraph{First-order equations:}
these arise from requiring the coefficients of $p_1$ and $p_2$ to be zero. Namely, we have
\begin{equation}
    B(x,y)k_2(x,y) + y\,\partial_y W(x,y) + \partial_x m_1(x,y) = 0,
\end{equation}
\begin{equation}
    B(x,y)k_1(x,y) - \partial_y m_1(x,y) + 2x\,\partial_y W(x,y) - y\,\partial_x W(x,y) = 0.
\end{equation}

\paragraph{ Zero-th order equation:}
this is found by imposing the zero order terms in the momenta to vanish. We obtain
\begin{equation}
k_1(x,y) \partial_x W(x,y)+k_2(x,y) \partial_y W(x,y)=0.\label{X10order}
\end{equation}
\subsection{Determining equations for $X_2$}\label{sec:eqX2}
Similarly, we can derive the determining equations for the integral $X_2$. They read as follows.
\paragraph{Second-order equations:}
\begin{equation}
    (b + x(d - 2cy))B(x,y) + \partial_x s_1(x,y) = 0, \label{eqs11}
\end{equation}
\begin{equation}
    (b + x(d - 2cy))B(x,y) - \partial_y s_2(x,y) = 0, \label{eqs12}
\end{equation}
\begin{equation}
    2(2a - cx^2 - dy + cy^2)B(x,y) - \partial_y s_1(x,y) - \partial_x s_2(x,y) = 0. \label{eqs13}
\end{equation}

\paragraph{First-order equations:}
\begin{align}
    &-B(x, y) s_2(x, y) + (b + dx - 2cxy) \partial_y W(x,y) - \partial_x m_2(x, y) \nonumber\\
    &+ 2(a - dy + cy^2) \partial_x W(x,y) = 0,
\end{align}
\begin{align}
   & B(x, y) s_1(x, y) - \partial_y m_2(x, y) + 2( cx^2- a) \partial_y W(x,y) \nonumber\\
   &+ (b + dx - 2cxy) \partial_x W(x,y) = 0.
\end{align}

\paragraph{ Zero-th order equation:}
\begin{equation}
s_1(x,y) \partial_x W(x,y)+s_2(x,y) \partial_y W(x,y)=0.\label{X20order}
\end{equation}
\subsection{Compatibility conditions}\label{sec:comp} 
As already observed, a brute force approach to solve the determining equations for the integrals would fail, due to the number of unknown functions and arbitrary constants in the integrals. This is why we look for compatibility conditions for the above equations. 
\paragraph{Compatibility conditions for the magnetic field $B$.} 
We assume that the functions $k_j$ and $s_j$ are smooth, $j=1,2$. Therefore, they must satisfy
\begin{equation}\label{Compk}
\partial_{xy}^2(\partial_y k_1(x,y) + \partial_x k_2(x,y))=\partial_{yy}^2\partial_x k_1+ \partial_{xx}^2\partial_y k_2
\end{equation}
and similarly,
\begin{equation}\label{Comps}
\partial_{xy}^2(\partial_y s_1(x,y) + \partial_x s_2(x,y))=\partial_{yy}^2\partial_x s_1+ \partial_{xx}^2\partial_y s_2\,.
\end{equation}
Let us take  the  second order equations for $X_1$, solve them for $\partial_x k_1$, $\partial_y k_1$  and  $\partial_y k_2$ and substitute into \eqref{Compk}. Similarly, we take the second order equations for $X_2$, solve them for $\partial_x s_1$, $\partial_y s_1$ and  $\partial_y s_2$  and substitute into  \eqref{Comps}.  In this way from \eqref{Compk} and \eqref{Comps} we obtain the following compatibility conditions for the magnetic field:
\begin{equation}\label{CompB1}
    4\,\partial_y B(x,y) + y\,\partial_{yy}^{2} B(x,y) +2x\,\partial_{xy}^{2}  B(x,y) - y\,\partial_{xx}^{2}  B(x,y) = 0 
\end{equation}
and
\begin{align}
    &-8cx\,\partial_y B(x,y) + (b + x(d - 2cy))\,\partial_{yy}^{2} B(x,y) - 4(d - 2cy)\,\partial_x B(x,y) \notag \\
    &\quad + 2\bigl(2a - dy + c(-x^2 + y^2)\bigr)\,\partial_{xy}^{2} B(x,y)  \notag \\ 
    & - (b + x(d - 2cy))\,\partial_{xx}^{2} B(x,y) = 0\,,\label{CompB2}
\end{align}
respectively.

We do not consider the case $b=d=c=0$, see remark 2, Section \ref{sec:strategy}.
Thus, we can multiply equation \eqref{CompB1} by $ y ^{-1}(b + d x - 2 c  x y)$ and add it to equation \eqref{CompB2}; then after multiplying by $y/2$, we obtain the simpler equation
\begin{eqnarray}
 &\left(2 a y-b x+c y \left(x^2+y^2\right)-d \left(x^2+y^2\right)\right) \partial_{xy}^2 B(x,y) \nonumber\\
& -2 (b+d x) \partial_y B(x,y)-2 y (d-2 c y) \partial_x B(x,y)=0\,.\label{CompB2s}
\end{eqnarray}
 In the following Sections \ref{sec:par-ell-Sol}-\ref{sec:acZero}  we shall start by solving conditions \eqref{CompB1}, \eqref{CompB2s} for the magnetic field. However, despite the fact that we managed to obtain equations in which the only unknown function is the magnetic field, depending on the constants $a,b,c,d$, we can not always find their solution, even with the help of computer algebra systems such as, e.g. Wolfram Mathematica. However, we can find additional compatibility conditions, managing a complete solution of the second order determining equations,  at least for special values of the constants $a,b,c,d$.\\
\paragraph{Compatibility conditions for $W$.}
We also assume $m_j$ to be  smooth functions, $j=1,2$. Therefore, mixed second partial derivatives of $m_j$ must commute, namely
\begin{equation}
\partial_{xy}^2 m_j= \partial_{yx}^2 m_j, \;\;j=1,2. \label{compm}
\end{equation}
By solving the first order determining equations for the partial derivatives of $m_1$ and $m_2$ and subsituting into \eqref{compm},  we find the conditions
\begin{align}\label{CompW1}
    &-k_2(x,y)\,\partial_y B(x,y)  - k_1(x,y)\,\partial_x B(x,y) \nonumber \\
    &\quad - B(x,y)(\partial_y k_2(x,y) + \partial_x k_1(x,y)) - 3\,\partial_y W(x,y) \nonumber\\
    &  - y\,\partial_{yy}^{2} W(x,y) - 2x\,\partial_{xy}^{2} W(x,y) + y\,\partial_{xx}^{2} W(x,y)=0
\end{align}
and
\begin{align}
    &-s_2(x, y)\,\partial_y B(x,y) - s_1(x, y)\,\partial_x B(x,y) - B(x,y)\,\partial_y s_2(x,y)  \notag \\
    &- B(x,y)\,\partial_x s_1(x,y) - 6cx\,\partial_y W(x,y) + (-3d + 6cy)\,\partial_x W(x,y) \nonumber \\
    & + (b + dx - 2cxy)\,\partial_{yy}^{2} W(x,y) - (b + dx - 2cxy)\,\partial_{xx}^{2} W(x,y)  \notag \\
    &+ (4a - 2cx^2 - 2dy + 2cy^2)\,\partial_{xy}^{2} W(x,y) = 0\,,\label{CompW2}
\end{align}
respectively. If \eqref{CompB1}, \eqref{CompB2}  and the second order equations are solved for $B$ and $k_j, s_j$, equations \eqref{CompW1}, \eqref{CompW2} give a compatibility condition for  the electrostatic potential $W$.\\
\paragraph{Compatibility condition for the $s_j$, $k_j$ functions.}
Finally, another compatibility condition arises from the zero-th order equations. First of all, let us notice that, excluding cases 1-2 of Section \ref{sec:strategy}, the functions $s_j$ and $k_j$ can not be identically zero, otherwise the second order equations would imply vanishing magnetic field. Thus, the zero-th order equation for both integrals can be solved for $\partial_y W(x,y)$, obtaining
$$\partial_y W(x,y)=-\frac{k_1(x,y) \partial_x W(x,y)}{k_2(x,y)}\,$$
and
$$\partial_y W(x,y)=-\frac{s_1(x,y) \partial_x W(x,y)}{s_2(x,y)}$$
for  $X_1$ and $X_2$, respectively. Therefore, we conclude that necessarily
\begin{equation}\label{EQksW}
\left(\frac{s_1(x,y)}{s_2(x,y)}-\frac{k_1(x,y)}{k_2(x,y)}\right) \partial_x W(x,y)=0\,.
\end{equation}
Unless  $\partial_x W(x,y)=0$, the above equation gives a compatibility condition for the $s_j$ and $k_j$ functions, namely
\begin{equation}\label{Compks}
k_2(x,y) s_1(x,y)-k_1(x,y) s_2(x,y)=0.
\end{equation}
As we shall see in the following, after solving the  second order determining equations, it is very effective to continue by considering equation \eqref{EQksW}, rather then proceeding with the first order equations, or their compatibility conditions, cf. \eqref{CompW1}, \eqref{CompW2}.
\section{Case $b=d=0$, $a,c$ both not zero: the integral $X_2$ is of elliptic type}\label{sec:par-ell-Sol}
Let us start by  solving the determining equations for the integrals in the case the integral $X_2$ is of elliptic type. Thus, we consider here $b=d=0$ and $a,c$ both not zero in \eqref{IntX2}. By scaling, we can set $c=1$, so to bring the integral into the standard form iv), see Section \ref{sec:goal}.

\subsection{Solution of the  second order equations}\label{sec:par-ell-2ndOrder-Sol}
Let us substitute $b=d=0, c=1$ into the compatibility condition \eqref{CompB2s} for the second order equations. We obtain
\begin{equation}\label{SimplCompB2Ell}
4y\partial_{x}B + (2a+x^2+y^2)\partial_{xy}^2B = 0.
\end{equation}
The above equation can be integrated w.r.t. y, yielding
\begin{equation}
\partial_x B(x,y)
= \frac{B_1(x)}{\bigl(2a + x^{2} + y^{2}\bigr)^{2}} \,,\label{SolBx}
\end{equation}
where $B_1(x)$ is an arbitrary function. By differentiating \eqref{CompB1} with respect to $x$, and substituting \eqref{SolBx} into it, we then obtain
$$\frac{y \left(\left(2 a+x^2+y^2\right)^2 B_1''(x)+48 a B_1(x)\right)}{\left(2 a+x^2+y^2\right)^4}=0\,.$$
Since the numerator of the above equation is polynomial in $y$, all coefficients of all powers of y, collected order by order, must be zero. This provides a set of differential equations for the function $B_1$. From the coefficient of $y^5$ we see that necessarily
$B_1''(x) = 0$, thus $B_1(x) = B_2 + B_3 x$, $B_2, B_3\in$ $\mathbb R$. However,
the equations coming from the lower order terms in $y$ imply that  $B_2 = B_3 = 0$. Therefore,
$B_1(x) = 0$, with the conclusion that $\partial_x B(x,y)=0$.
Thus, $B$ is function only of $y$, i.e. $B(x, y) = B(y)$. By substituting this into
 \eqref{CompB1}, we arrive at the following equation for $B(y)$
\begin{equation}
    4 B'(y) +  B'' (y) y=0 \,,
\end{equation}
whose solution leads to
\begin{equation}
B(x, y)=-\frac{\beta_2}{3y^3}+\beta_1, \;\; \beta_1,\beta_2\in\mathbb R\,. \label{B}
\end{equation}
Now let us proceed by substituting \eqref{B} into the second order equations and solve them for $k_1$, $k_2$, $s_1$ and $s_2$. Their solution is now straightforward and reads
\begin{eqnarray}
k_1(x,y)&=&-\frac{\beta_2x}{3 y^2}+\beta_1 x y+k_{11} y+ k_{12},\label{Solk1ell}\\
k_2(x,y)&=& -\frac{1}{2} \beta_1 ( 3x^2+y^2)-\frac{\beta_2}{3 y}-k_{11} x+ k_{21},
\end{eqnarray}
\begin{eqnarray}
s_1(x,y)&=&\frac{2 a\beta_2}{3 y^2}+x^2 \left(\beta_1 y-\frac{\beta_2}{3 y^2}\right)+\beta_1 y^3+s_{11} y+s_{12},\\
s_2(x,y)&=&x \left(4 a\beta_1- \beta_1 y^2-\frac{2 \beta_2}{3 y}-s_{11}\right)-\beta_1 x^3+s_{22},\label{Sols2ell}
\end{eqnarray}
where $k_{11}, k_{12}, k_{21}, s_{11}, s_{12}, s_{22} \in\mathbb R$. 

\subsection{Solution of the compatibility conditions for the zero-th order equation}\label{sec:par-ell-ZerothOrder}
Let us substitute the solutions \eqref{Solk1ell}-\eqref{Sols2ell} for $k_1$, $k_2$, $s_1$ and $s_2$ into equation \eqref{EQksW}. We obtain
$$ P(x,y) \partial_x W(x,y) =0\,, $$
where
\begin{equation}
\begin{aligned}
& P(x,y) = 2 a\beta_2 -  (\beta_2 - 3 \beta_1 y^{3})x^{2} + 3  (\beta_1 y^{3}   +  s_{13}y + s_{12})y^{2} \\
&+ \frac{2 (-\beta_2 x + 3 (k_{12} +  (k_{11} + \beta_1 x)y) y^{2})}{6 k_{21} y - 6 k_{11} x y - 2 \beta_2 - 3 y (3 x^{2} + y^{2}) \beta_1} \left(2  \beta_2 x \right. \\
& \left.+ 3 x y (\beta_1(-4 a + x^{2} + y^{2})  + s_{13}) - 3  s_{22} y\right)\,. \label{PolEll}
\end{aligned}
\end{equation}
Thus, there are two possibilities :
\begin{enumerate}
\item $\partial_x W(x,y)=0$. However, in this case $W$ would be only function of $y$. Since also the magnetic field does not depend on the $x$ variable, then necessarily in proper gauge choice,  $p_1$ is an integral. Therefore, the system would have one Cartesian type integral and this case was studied in \cite{BeWin}.
\item $P(x,y)=0$.
In this case,  we can reduce the  expression \eqref{PolEll} to common denominator and by imposing that it equals zero, we obtain an equation polynomial in $x,y$ that can be satisfied only if all 
coefficients of each monomial term in $x,y$ vanishes, order by order. From this it follows that $\beta_1=\beta_2=0$ and therefore the magnetic field vanishes. Thus, we conclude that in this case for non-vanishing magnetic field nothing of interest can be found.
\end{enumerate}

\section{Case $a=b=c=0$: the integral $X_2$ is of parabolic type}\label{sec:Par-Par}
Let us consider here the case $a=b=c=0$. Then by scaling, we can set $d=1$. Thus, also the integral $X_2$ is of parabolic type, see iii), Section \ref{sec:goal}. 
As in the previous case, let us start 
from the compatibility condition \eqref{CompB2s}. By setting $a=b=c=0$, $d=1$,  we obtain
\begin{equation}
     2x\, \partial_y B(x,y)
+ 2y\, \partial_x B(x,y)
+ (x^2 + y^2)\, \partial_{xy}^2 B(x,y)= 0.
\end{equation}
The above equation can be solved for the magnetic field. We find
\begin{equation}
    B(x,y) = \frac{B_1(x)+ B_2(y)}{\left(x^2+y^2\right)}\,,\label{SolBParCompB2}
\end{equation}
where $B_1(x)$, $B_2(y)$ are arbitrary functions. By substituting \eqref{SolBParCompB2} into \eqref{CompB1}, we obtain an equation for the functions $B_1(x)$ and $B_2(y)$, namely
$$ B_1''(x)- B_2''(y)=0\,. $$
The above equation implies $B_1''(x)= B_2''(y)=\beta_0$, $\beta_0\in\mathbb R$. Thus,
\begin{equation}
    B(x,y) = \frac{ \beta_0}{2} + \frac{\beta_{11} + x\,\beta_{12} + y\,\beta_{22}}{x^2 + y^2} \,,\label{SolBPar}
\end{equation}
where $\beta_0, \beta_{ij}$ are constants.

Once we plug \eqref{SolBPar} into the second order equations   for $X_1$ and $X_2$, they can be easily solved. We obtain

\begin{flalign*}
k_1(x, y) &= -\frac{k_{11} y}{2} + (\beta_{11} +  \beta_{22}y)\arctan\left(\frac{x}{y}\right)  + k_{12}  \notag \\
&+ \frac{1}{2}  (\beta_0 x + \beta_{12} \ln(x^2 + y^2))y\,,\\
k_2(x,y) &= -\frac{1}{4} \Big( -2 k_{11} x + 3 \beta_0 x^2 + \beta_0 y^2 + 4  \beta_{12}x + 4 \beta_{22} y- 4 k_{22}  \\
&+ 4  \beta_{22} x \arctan\left(\frac{x}{y}\right) + 2 (\beta_{11} + x \beta_{12} ) \ln(x^2 + y^2) \Big) 
\end{flalign*}

and
\begin{align*}
s_1(x,y) &= \frac{1}{4}\Bigl(
   -4\beta_{12} y \arctan\!\left(\frac{y}{x}\right) 
   -  (\beta_0 x + 4 \beta_{12})x\, \\
   &- 2 \bigl(\beta_{11}  +  \beta_{22}y\bigr)
     \ln\left(x^2 + y^2\right)
\Bigr)-s_{11}y -\frac34 \beta_0 y^2- \beta_{22}y+ s_{12}, \\
s_2(x,y) &= \frac{\beta_0 x y}{2}
+ \bigl(\beta_{11} +  \beta_{12}x \bigr)\arctan\!\left(\frac{y}{x}\right)
   \\
&
+ \frac{1}{2}  \beta_{22}x\,
\ln\!\left(x^2 + y^2\right)
+s_{11} x + s_{22}\,.
\end{align*}


We are looking for globally defined integrals, therefore we have to exclude from  $s_j$ and $k_j$ terms that have domain of definition smaller than the one of the magnetic field \eqref{SolBPar}. This implies $\beta_{11}=\beta_{12}=\beta_{22}=0$.
By substituting into \eqref{SolBPar}, we find that the magnetic field is constant.


By proceeding as in the previous Section,  we then substitute the solution so found for $k_j$ $s_j$  and $B$ into \eqref{EQksW}
and arrive again to the conclusion that either $\beta_0 = 0$ and therefore $B(x,y) = 0$, or $W(x,y)$ is constant.

\section{$c=0$, $d\neq0$ and $a$ or $b$ not zero: the integral $X_2$ is of ``non-standard'' parabolic type}\label{sec:nonstandardPar}
Let us consider here the case in which $c=0$, $d\neq0$ and also $a$ or $b$ (or both) is not zero in the integral \eqref{IntX2}. Thus, here we study the case in which the integral $X_2$ is of non-standard parabolic type. We have to distinguish several sub-cases, depending on the value of $a$ and $b$, that are investigated in the following Subsections.

\subsection{Case $a=c=0$, $b,d\neq0$}\label{sec:acZero}
As in the previous Section, we start from the compatibility condition \eqref{CompB2s}. Here we consider $a=c=0$, but $b\neq0$. In this case, we cannot find an explicit solution for equation \eqref{CompB2s}. However, equation \eqref{CompB2s} can be integrated with respect to $y$, yielding
\begin{equation}
    2\,(b + d x)\,B(x,y) - B_1(x) + (b x + d x^2 + d y^2)\,\partial_x B(x,y) = 0\,,
    \label{CompB2Inta0N}
\end{equation}
where $B_1(x)$ is an arbitrary function arising from the integration. Afterwards, we can proceed as follows.
By substituting the expression for $\partial_x B$ obtained by solving \eqref{CompB2Inta0N}, namely ($b$, $d$ are both non-zero by assumption)
\begin{equation}
\partial_x B(x,y)= \frac{-2 b B(x,y)-2 d x B(x,y)+B_1(x)}{b x+d x^2+d y^2} \,, \label{SubBx}
\end{equation}
into \eqref{CompB1}, after multiplying by $\frac{(b x + d(x^2 + y^2))^2}{y}$,
we arrive at the equation
\begin{equation}
\begin{aligned}
&(6 b^{2} + 4 b d x - 2 d^{2} (x^{2} + y^{2}))\, B(x, y)
- 3 b\, B_1(x) \\
&\quad - (b x + d (x^{2} + y^{2}))
\bigl(-B_1'(x) + 4 d y\, \partial_y B(x, y) \\
&+ (b x + d (x^{2} + y^{2}))\, \partial^2_{yy} B(x, y)\bigr) = 0\,.
\end{aligned}
\label{CompB1a0}
\end{equation}
Next, we  multiply  \eqref{CompB1a0} by $( b x+d \left(x^2+y^2\right))^{-1}$, differentiate it with respect to $x$ and substitute \eqref{SubBx} into it. This leads to
\begin{align}
&-6 b B(x,y) \left(3 b^2+6 b d x+d^2 \left(3 x^2-y^2\right)\right) \nonumber \\
&+\left(b x+d (x^2+y^2)\right)\left( (b x+d (x^2+y^2) )\left( \partial^2_{yy}B(x,y)
+B_1''(x)\right)\right.\nonumber \\
&\left.-3 b B_1'(x)\right) +3 b (3 b+4 d x) B_1(x)=0.
\label{CompB1a00}
\end{align}
 By solving the above equation for $\partial^2_{yy} B(x,y)$ (here we assume $b\neq0$) and substituting back into \eqref{CompB1a0} we obtain
\begin{eqnarray}
&& -4 b B(x,y) \left(3 b^2+8 b d x+d^2 \left(5 x^2-y^2\right)\right) \nonumber\\
&&\left. + (b x+d (x^2+y^2)\right) \left(-4 b d y \partial_y B(x,y) \right. \nonumber\\
&& \left. +B_1''(x) \left(b x+d (x^2+y^2)\right)-2 b B_1'(x)\right)\nonumber\\
&&+6 b (b+2 d x) B_1(x)=0\,.\label{eqBya0}
\end{eqnarray}
Afterwards, we solve equation \eqref{eqBya0} for $\partial_y B(x,y)$. In this way, together with \eqref{SubBx}, we have explicit expressions for both partial derivatives of $B$ where no higher order derivative for $B$ is involved. Then, by requiring that the mixed partial derivatives of $B$ are equal, i.e.
\begin{equation}
\frac{\partial^2 B}{\partial x \partial y}
-
\frac{\partial^2 B}{\partial y \partial x}
= 0,
\end{equation}
from the expression we found for $\partial_x B(x,y)$ and $\partial_y B(x,y)$ we obtain an equation where the only unknown functions are $B_1$, with its derivatives (up to third order),  and $B$, that instead appears without any of its derivatives. Thus, finally we have an equation that can be solved for $B$, yielding 
\begin{equation}
\begin{aligned}
B(x,y) &= \frac{1}{4b \Big(3 b^3 + 6 b^2 d x - 16 d^3 x y^2 + b d^2 (3 x^2 - 13 y^2)\Big)} \cdot\\
& \quad  \Bigg(  2b \Big(3b^2 + 4b d x + 4d^2(x^2 - 3y^2)\Big) B_1(x) \\
& \quad - \Big(bx + d(x^2 + y^2)\Big) 
\Big( 
    2b(b + 4 d x) B_1'(x) \\
& + \Big(bx + d(x^2 + y^2)\Big) 
    \Big( (b + 4 d x) B_1''(x) 
    + \Big(bx \\
    &+ d(x^2 + y^2)\Big) B_1'''(x) \Big) 
\Big) 
\Bigg)\,.
\end{aligned}
\end{equation}
We substitute this expression for $B(x,y)$ back into the original compatibility condition \eqref{CompB2Inta0N}, obtaining an expression polynomial in $y$ with coefficients depending on $x$, the function $B_1(x)$ and its derivatives. Requiring that the coefficients of all powers of $y$, collected order by order, vanish, we obtain the following solution for $B_1(x)$
$$B_1(x)=\frac{\beta_0 d x}{b}+\beta_0, \,\, \beta_0\in\mathbb R\,,$$
that implies
\begin{equation}
  B(x,y) = \frac{\beta_0}{2b}\,.
  \end{equation}
Thus, the function $B(x,y)$ reduces to a constant. The solution for the determining equations for the integrals follows then as in Section \ref{sec:ConstB}.

\subsection{Case $b=c=0$, $a, d\neq0$}\label{sec:bcZero}
By substituting $b=c=0$ into equation \eqref{CompB2s}, we obtain
\begin{equation}
    2 d x\, \partial_y B(x,y)
+2 d y\, \partial_x B(x,y)
- (2 a y - d (x^2 + y^2))\, \partial_{xy}^2 B(x,y) = 0\,. 
\end{equation}
As previously, we are not able to find explicit solution of the above equation. However, we can integrate it  with respect to $x$, finding
\begin{equation}
    2 d y\, B(x,y)
- B_1(y)
+ (d x^2 + d y^2 - 2 a y )\, \partial_y B(x,y) =0\,.\label{CompB2Intb0}
\end{equation}
Then proceeding similarly as in the previous Section, we solve the above equation for  $\partial_yB(x,y)$ and substitute the so found solution into \eqref{CompB1}, arriving at
\begin{align}
& (d (x^2+y^2)-2 a y) \partial^2_{xx} B(x,y)+4 d x \partial_x B(x,y)-B_1'(y)\nonumber \\
&- (d(x^2+y^2)-2 a y) ^{-1} (2 d \left(d \left(x^2+y^2\right)  B(x,y)-8 a y\right)+6 a B_1(y))=0 \,,\label{eqCompB1Bxx}
\end{align}
where we also multiplied \eqref{CompB1} by $\frac{\left(d \left(x^2+y^2\right)-2 a y\right)}{y}$.
We differentiate \eqref{eqCompB1Bxx} with respect to $y$ and after substituting into it the expression for $\partial_y B(x,y)$ found by solving \eqref{CompB2Intb0}, we arrive at
\begin{align*}
& (d (x^2+y^2) - 2 a y) 
(
    (d (x^2+y^2) - 2 a y)(2 a \, \partial_{xx}^2 B(x,y) + B_1^{''}(y)) 
    - 6 a \, B_1'(y))  \\
& +12 a d^2 (x^2 - 3 y^2) B(x,y) 
- 12 a (a - 2 d y) B_1(y) = 0 \,.
\end{align*}
We solve the above equation for $\partial_{xx}^2 B(x,y)$ and substitute it into \eqref{eqCompB1Bxx}, obtaining an equation where only $\partial B_x$, $B$, $B_1$ and its derivatives are present, namely
\begin{align}
&\left(d \left(x^2+y^2\right)-2 a y\right) \left(-8 a d x \partial_x B(x,y)+B_1''(y) \left(d \left(x^2+y^2\right)-2 a y\right) -4 aB_1'(y)\right)\nonumber\\
&  +8 a d B(x,y) \left(4 a y+d \left(x^2-5 y^2\right)\right)-24 a (a-d y) B_1(y)=0. 
\end{align}
Finally, similarly to the previous Section, by solving \eqref{CompB2Intb0} for $\partial_y B$ and by solving the above equation for $\partial_x B$ and then  by 
 imposing that the mixed derivatives of $B$ are equal, we  arrive at the following equation for $B$
 \begin{align*}
& (d (x^2+y^2) - 2 a y ) 
\bigg((d (x^2+y^2) - 2 a y ) \bigg( 
        (d (x^2+y^2) - 2 a y)  B_1'''(y)
         \\
& +  (4 d y - 6 a)B_1''(y)\bigg) - 8 a (3 a - 2 d y) B_1'(y)\bigg)
\\
&+16 a d^2 B(x,y) (3 a (x^2+y^2) - 8 d x^2 y)  \nonumber\\
&- 16 a B_1(y) (3 a^2 - 2 a d y + d^2 (y^2 - 3 x^2)) = 0 \,.
\end{align*}
We solve the above equation for $B(x,y)$ and substitute the so found solution back into \eqref{CompB2Intb0}. In this way, we obtain an equation polynomial in $x$. By imposing that  all coefficients of all powers of $x$ are zero, order by order, we obtain a set of differential equations for the function $B_1$ whose solution leads to the conclusion that the magnetic field must be constant. The solution of the determining equations for the integrals then follows as in Section \ref{sec:ConstB}.

\subsection{Case $c=0$, $a,b,d \neq 0$}\label{sec:cZero}

By setting $c = 0$ into equation \eqref{CompB2s}, we obtain 
\begin{eqnarray}
&&  -2 (b + d x) \partial_{y} B(x,y) - 2 d y \partial_{x} B(x,y) \nonumber\\
 &&   + \left( -b x + 2 a y - d (x^2 + y^2) \right) \partial_{xy}^2 B(x,y) = 0\label{CompB2sc0N}
\end{eqnarray}
In this case, all remaining constants $a,b,d$ are not zero and the above equation cannot be explicitly integrated. However, we can proceed as follows. Let us substitute
\begin{equation}
k_1(x,y)= y K_1(x,y)\label{subK1}
\end{equation}
into  \eqref{eqk11}. Then by differentiating  \eqref{eqk11} w.r.t. $y$, we find that
\begin{equation}
\partial_y B(x,y)-\partial_{xy}^2K_1(x,y)=0\,.
\end{equation}
By using this, we rewrite   \eqref{CompB2sc0N} as
\begin{eqnarray}
&&\left(2 a y-b x-d \left(x^2+y^2\right)\right)\partial_{xy}^2B(x,y) -(b+2 d x) \partial_y B(x,y)\nonumber\\
&&-b \partial_{xy}^2 K_1(x,y)-2 d y \partial_x B(x,y)=0\,.
\end{eqnarray}
By integrating the above expression w.r.t. $x$,  we obtain that ( $b\neq0$ by assumption)
\begin{equation}
    \partial_{y}K_1(x, y) = \frac{-2 d y B(x, y) + B_2(y) - \left( b x - 2 a y + d (x^2 + y^2) \right) \partial_{y}B(x, y)}{b}\,,\label{SubK1y}
\end{equation}
where $B_2(y)$ is an arbitrary function arising from the integration.

Afterwards, we differentiate \eqref{eqk12} w.r.t. $x$ and we substitute into  the expressions for $\partial_x k_2$  that we can obtain from \eqref{eqk13}, then we also substitute \eqref{subK1}, \eqref{SubK1y} . This yields an additional compatibility condition for $B$, namely

\begin{equation}
\begin{split}
& -6 d y B(x, y) + 2 B_2(y) - 2 d x^2 \partial_{y}B(x, y) \\
& - y \left( -B_2'(y) - 6 (a - d y) \partial_{y}B(x, y) \right. \\
& + \left( b x - 2 a y + d (x^2 + y^2) \right) \partial_{yy}^2B(x, y) \\
& \left. + b \partial_{x}B(x, y) \right) =0\,.
\end{split}
\label{CompB3}
\end{equation}
Similarly, to be able to integrate \eqref{CompB2sc0N} w.r.t. y explicitly, we substitute 
\begin{equation}\label{subS2}
s_2(x,y)= (b+d x) S_2(x,y)
\end{equation}
into \eqref{eqs12}. In this way, by differentiating \eqref{eqs12} w.r.t. $x$, we find that
$\partial_{x}B = \partial_{xy}^2S_2(x,y)$.  This allows to rewrite  \eqref{CompB2sc0N} as
\begin{eqnarray}
&&  -2 (b + d x) \partial_{y} B(x,y) - 2 d y \partial_{x} B(x,y) \nonumber\\
 &&   + \left( -b x + 2 a y - d (x^2 + y^2) \right) \partial_{xy}^2 B(x,y) + 2 a\partial_x B(x,y)\nonumber\\
 && -2a\partial_{xy}^2S_2(x,y)= 0\,.
\end{eqnarray}
The above expression can now be integrate w.r.t. y and we get ($a\neq0$ by assumption here)
\begin{equation}
     \partial_{x}S_2(x, y) = \frac{-2 (b + d x) B(x, y) + B_1(x) - \left( b x - 2 a y + d (x^2 + y^2) \right) \partial_{x}B(x, y)}{2 a}\,,\label{SubS2x}
\end{equation}
where $B_1(x)$ is an arbitrary function.
Proceeding as before, we differentiate equation \eqref{eqs13} w.r.t. $x$, we substitute into it the expression for $\partial_x s_1$ that can be derived from \eqref{eqs11}. Then we also substitute   \eqref{subS2} and  \eqref{SubS2x} into it.  This leads to another compatibility condition for $B$:
\begin{equation}
\begin{split}
& (b + d x) \partial_{y}B(x, y) + 2 (2 a - d y) \partial_{x}B(x, y) \\
& + \frac{d \left( 2 (b + d x) B(x, y) - B_1(x) + (x (b + d x) - 2 a y + d y^2) \partial_{x}B(x, y) \right)}{a} \\
& + \frac{(b + d x) \left( 2 d B(x, y) - B_1'(x) + (3 b + 4 d x) \partial_{x}B(x, y) \right.}{2 a} \\
& + \frac{\left. (x (b + d x) - 2 a y + d y^2) \partial_{xx}^2B(x, y) \right)}{2 a} = 0\,.
\end{split}
\label{CompB4}
\end{equation}
Now, we have four compatibility conditions for $B$, namely equations \eqref{CompB1}, \eqref{CompB2sc0N}, \eqref{CompB3} and \eqref{CompB4}.
We solve \eqref{CompB2sc0N} for $\partial_{xy}^2B(x,y)$, \eqref{CompB3} for $\partial_{yy}^2B(x,y)$ and \eqref{CompB4} for $\partial_{xx}^2B(x,y)$ and substituting  into compatibility condition \eqref{CompB1}, we arrive at
\begin{equation}
\begin{split}
& 2 d y B_1(x) + (b + d x) \left( -2 B_2(y) + y (B_1'(x) - B_2'(y)) \right. \\
& \left. + 2 d (x^2 + y^2) \partial_{y}B(x, y) \right) - 2 y \left( (b + d x)^2 + (-2 a + d y)^2 \right) \partial_{x}B(x, y) = 0 \,.\label{CompB1s}
\end{split}
\end{equation}
We  differentiate this equation w.r.t $x$  and substitute into it  the expressions for  $\partial_{xy}^2B(x,y)$, $\partial_{yy}^2B(x,y)$, $\partial_{xx}^2B(x,y)$ found from \eqref{CompB2sc0N}, \eqref{CompB3} and \eqref{CompB4}, respectively. Then we also substitute the solution for $\partial_y B$ that we can find from \eqref {CompB1s}.
In this way we obtain an equation which depends on $B$ only through  $\partial_{x}B(x,y)$ and $B$. Similarly, by differentiation \eqref{CompB1s} w.r.t. $y$ we arrive at an equation which depends on  $B$ only through $B$ and $\partial_{y}B(x,y)$. We solve these equations for $\partial_{x}B(x,y)$ and $\partial_{y}B(x,y)$ and substitute the so found expressions for $\partial_{x}B(x,y)$ and $\partial_{y}B(x,y)$ into \eqref{CompB1s}. In this way we obtain an equation which contains only $B(x,y)$, $B_1(x)$, $B_2(y)$ and the derivatives of $B_1(x)$ and $B_2(y)$. This equation can be solved for $B(x,y)$. Here the use of algebraic manipulators such as Mathematica becomes fundamental. 

The expression for $B(x,y)$ so found  must also satisfy the original compatibility condition \eqref{CompB2sc0N}.  By  substituting it  into \eqref{CompB2sc0N},  we arrive at an equation for $B_1(x)$ and $B_2(y)$,  where the coefficients of $B_2$ and its derivatives are polynomial in $x$ (similarly, the coefficients of $B_1$ and its derivatives are polynomial in $y$). Thus, by repeatedly differentiating this equation with respect to $x$, we can eliminate the dependence on $B_2(y)$ and its derivatives and we are left with a polynomial equation in $y$ with coefficients depending on $B_1(x)$ and its derivatives. By imposing that the coefficients of each power of $y$ vanish, we find that
\begin{equation}
B_1(x) = \frac{\beta_{11}}{(b + d x)^2}   + \beta_{12} + \beta_{13} x\,,\; \beta_{ij}\in\mathbb R\,.
\end{equation}
Similarly, by differentiating w.r.t. $y$, we can eliminate the dependence on $B_1(x)$ and its derivative, obtaining 
\begin{equation}
B_2(y) = \beta_{13}y - \frac{\beta_{21}}{2y^2} + \beta_{22}\,,\; \beta_{ij}\in\mathbb R\,.
\end{equation}
Finally, substituting these solutions for $B_1(x)$ and $B_2(y)$ back into the original equation, we obtain that necessarily $\beta_{22}=0$ and $b \beta_{13}-d \beta_{12}=0.$

Afterwards, the solution for $B$
simplifies to
\begin{equation}
    B(x,y) = \frac{\beta_{12}}{2b}\,.
\end{equation}
Thus, also in this case,  we found that the magnetic field must be constant.

\section{Solution for constant magnetic field}\label{sec:ConstB}
In both computations of Sections \ref{sec:acZero}, \ref{sec:cZero} the compatibility conditions \eqref{CompB1}, \eqref{CompB2s} imply that the magnetic field is constant. Actually, the constant magnetic field is trivially a solution for equations \eqref{CompB1}, \eqref{CompB2s} (though not the only one, see \eqref{B}, \eqref{SolBPar}). Thus, let us assume
that the magnetic field is constant, namely
\begin{equation}
B(x,y) = \beta\,, \;\;\beta\in\mathbb R\,,\label{SolBConst}
\end{equation}
and let us proceed in the solution of the second order determining equations for the integrals.
Under the assumption \eqref{SolBConst}, the second-order determining equations significantly simplify.
Their solution for $k_i$, $s_j$ is straightforward and reads
\begin{equation}
    k_1(x,y) = k_{11}\,y + \beta xy + k_{12}\,,
\end{equation}
\begin{equation}
k_2(x,y) = -k_{11}\,x - \frac{3}{2} \beta x^2 - \frac{1}{2} \beta y^2 + k_{22}\,,
\end{equation}
\begin{equation}
    s_1(x,y) = -s_{11}\,y - \beta (  b x + \frac{1}{2} d x^2 -4a  y
- c  x^2 y + \frac{3}{2}  d y^2- c y^3)+ s_{12}\,, 
\end{equation}
\begin{equation}
s_2(x,y) = s_{11}\,x + \beta(b y + d x y - c x(x^2 + y^2)) + s_{22}\,,
\end{equation}
where $s_{ij}, k_{ij} \in \mathbb R$.
By proceeding as in Section \ref{sec:par-ell-ZerothOrder}, we substitute the so found solution for $k_i$, $s_j$ into the compatibility condition \eqref{EQksW} and conclude that either $\partial_y W=0$ or the  following polynomial in $x$ and $y$ must vanish:
\begin{equation}
\begin{aligned}
& \beta^2 \left( \frac{3}{4} d y^4 - \frac{1}{2} c y^5 
+ x^4\left(\frac{3}{4} d  - \frac{1}{2} c y \right)\right)  \\
&+ \beta  \left(\frac{1}{2} d k_{11} + \frac{3}{2} b \beta + c k_{12}\right)x^3  \\
&+  \beta\left(\frac{1}{2} s_{11} - 2 a \beta + c k_{22}\right) y^3
+ k_{22} s_{12}  \\
&+  \beta\Big(b  k_{11}  + \frac{3}{2} d y^2 \beta - c y^3 \beta
- \frac{1}{2} d  k_{22}  \\
&+ y\left(\frac{1}{2} s_{11}  - 6 a \beta + c k_{22}\right)
- \frac{3}{2} \beta s_{12}\Big) x^2 \\
&+ \beta \left(- b k_{11}  - \frac{3}{2} d  k_{22} - \frac{1}{2}  s_{12}\right)y^2
- k_{12} s_{22}  \\
&+ \left(- b \beta k_{12} - s_{11} k_{22} + 4 a \beta k_{22} - k_{11} s_{22}\right) y  \\
&+ \Big(- s_{11} k_{12}
+ y^2\left(\frac{1}{2} d k_{11} \beta - \frac{1}{2} b \beta^2 + c \beta k_{12}\right)  \\
&- b \beta k_{22} - k_{11} s_{12}
+ y\left(- 4 a k_{11} \beta - d \beta k_{12} - \beta s_{22}\right)\Big) x = 0.
\end{aligned}
\end{equation}
The requirement that all coefficients of all monomial terms in the above expression vanish (for non-zero $\beta$) leads to the conclusion that $a=b=d=c = 0$, $s_{12} = s_{22} = 0$ and $\frac{s_{11}\beta}{2} - 2a\beta = 0$. This implies $s_1(x,y)=0$ and $s_2(x,y)=0$.

This means that the second integral is trivially $X_2 = 0$, for non-vanishing constant $B$. Therefore, we conclude that the only possibility for superintegrabilty, with constant magnetic field, is $\partial_y W=0$, which implies that, in proper gauge choice, $p_1$ is an integral and thus, the system admits a Cartesian type integral. Therefore, the only possibility is for the potential to be constant \cite{BeWin}.

\section{Conclusions and perspectives}\label{Sec:conclusions}
In this work, we contribute to the classification of 2D quadratically superintegrable systems in a magnetic field.
So far, all results known in the literature show that, for a non-vanishing magnetic field, the only 2D system that admits integrals at most quadratic in the momenta is the CMF system. It remains open the question whether this is the only 2D quadratically superintegrable system with a non-vanishing magnetic field. 
To shed some light on the answer to this question, we focus here on the case in which the system \eqref{HamiltonianW} has two quadratic integrals, and one of them is of parabolic type. To our knowledge, this case has not yet been investigated  in the literature. We derived the compatibility conditions that the magnetic field  has to satisfy for such a pair of integrals to exist; see equations \eqref{CompB1}, \eqref{CompB2}.  Clearly, a constant magnetic field is a solution of the compatibility conditions. The computation in Section \ref{sec:ConstB} shows that, as a consequence of the determining equations for the integrals, in this case also the electrostatic potential must be constant. Thus, in case the magnetic field is assumed constant, there can be no result of interest.

By making no restriction on the form of the magnetic field, we could find explicit solutions  in several special cases, namely the cases in which the integral $X_2$ is of elliptic or of (non-standard) parabolic type. In both cases with elliptic and (standard) parabolic type integrals, we found that  the compatibility conditions for the magnetic field are solved by a non-constant magnetic field, see \eqref{B} and \eqref{SolBPar}, respectively. However, from the determining equations of the integrals it follows that the magnetic field then reduces to a constant one. In all other cases in which we could find  explicit solution for the compatibility conditions, we always found that the only possible solution is given by the constant magnetic field. This then implies that the electrostatic potential must also be constant.

Therefore, our computation seems to confirm that, in the presence of a magnetic field,  quadratic superintegrability  in the 2D Euclidean space is possible only if both the magnetic field and the electrostatic potential are constant. The present manuscript is a first step in a research plan that aims to confirm or disprove this conjecture.  It is in our plans to address the problem when the integral $X_2$ could not reduce to parabolic type by Euclidean transformations (i.e. when $c\neq0$ in \eqref{sec:eqX2}) and the case 
 in which the integral $X_1$ is taken of elliptic type. 
 
Furthermore, by allowing more general forms of the integrals, we could obtain richer results. Due to the presence of the velocity-dependent terms in the Hamiltonian, the investigation of the existence conditions  for integrals that are  rational or transcendental integrals looks promising \cite{HietarintaReview}.

Besides the direct approach to the solution of the determining equations that we followed in this work, other methods could  be employed to discover new (super)integrable systems, such as canonical transformations \cite{MS2, HietarintaReview} or coupling constant metamorphosis \cite{Ramani1984,KalninsStackelTr, Marchesiello2025}. See \cite{HietarintaReview} for a comprehensive review.

We considered here only the classical problem. In the corresponding quantum case the zero-th order determining equations for the integrals would contain quantum corrections \cite{BeWin, MSW}. Therefore, a priori we cannot exclude the existence of purely quantum systems, e.g. systems that would reduce to free motion in the classical limit.
It would be worth investigating whether the quantum corrections would lead to more interesting solutions, such as purely quantum systems with non-constant magnetic field.

\subsection*{Acknowledgements}

This work was supported by the ``Student Summer Research Program 2025 of FIT CTU in Prague''.

Computations in this paper were performed using computer algebra software  Mathematica\texttrademark\, by Wolfram Research, Inc. Champaign, IL.

\bibliography{mag-field}\label{lastpage}
\bibliographystyle{plain}

%
%
%
%
%
%
\end{document}